# Educating future engineers about LLMs: A scalable workshop

**March 2026**

R. Zhang, J. C. F. de Winter*, T. Dicke, D. Dodou, & Y. B. Eisma

*Faculty of Mechanical Engineering, Delft University of Technology, Delft, The Netherlands*

*Corresponding author: j.c.f.dewinter@tudelft.nl

**Abstract**

As large language models (LLMs) are increasingly integrated into engineering workflows, students require hands-on experience to learn how to collaborate with them critically. This paper presents a scalable gamified workshop designed for engineering Master's students to practice human-AI collaboration in navigation planning. Using a mobile web interface, across 10 workshop sessions, a total of 226 students wrote prompts for a non-reasoning and a reasoning LLM to solve grid-based navigation tasks of increasing complexity. The system returned robot-executable plans, trajectory visualizations, and automated scoring, culminating in a live demonstration on a Boston Dynamics Spot robot. In a post-workshop questionnaire, 81.5% reported substantial learning and 91.0% reported high engagement. Analysis of the submitted prompts revealed that students changed their strategies from step-by-step instructions for the non-reasoning LLM toward providing higher-level guidance for complex problem-solving tasks. We conclude that such interactive simulation-to-reality environments are viable for teaching the verification and collaboration skills necessary for responsible LLM use in engineering. Code is available at: https://github.com/renchizhhhh/LLM-robotics-workshop

## Introduction

Large language models (LLMs) are rapidly becoming part of how engineers write code, explore design alternatives, and communicate technical ideas. These developments create a dilemma for education. LLMs can be misused as a shortcut for completing assignments (Cotton et al., 2024), yet they also enable new forms of interactive learning, especially when students learn to treat the model as a collaborator whose output must be interpreted, tested, and verified (Kasneci et al., 2023; Lo, 2023). For engineering programs, the key question is no longer whether students will use LLMs, but how we can teach them to use these systems critically and responsibly.

Robotics provides a particularly concrete setting for this challenge because LLM outputs can be operationalized: a model's response can be turned into executable actions, and failures become immediately visible. Recent work has shown that LLMs can support robotics workflows in multiple ways, for example by mapping natural language to robot skills (Ahn et al., 2023; Black et al., 2025; Shridhar et al., 2022), generating code-based policies (Arenas et al., 2024; Liang et al., 2023), or producing symbolic plans that downstream systems can follow (Chen et al., 2024; Huang et al., 2022; Silver et al., 2024; Zhou et al., 2024). However, LLMs remain probabilistic and can produce plausible-looking outputs that violate constraints or fail in subtle ways (Kumar & Cohen, 2026). In safety-relevant domains such as robotics, this makes verification a core competence: users need to learn to anticipate failure modes, prompt the model appropriately, and validate the resulting plan before execution.

A practical complication is that 'LLM capability' is not monolithic. In robotics planning, non-reasoning models typically behave like pattern matchers: they can follow highly structured instructions but may fail when the problem requires multi-step constraint satisfaction (Kambhampati et al., 2024; Valmeekam et al., 2023). Newer reasoning LLMs can handle more complex planning-like tasks by performing internal reasoning, yet their performance still degrades as the search space grows (Shojaee et al., 2025; Zhang et al., 2026). Importantly, this difference is not only about task success, but also about how cognition is distributed between human and model. With non-reasoning models, a larger share of the "thinking" may have to be externalized by the user: successful prompting may require providing domain knowledge up front (e.g., decomposing into subgoals, or giving explicit waypoints). Reasoning models can be expected to complete simple tasks with less human guidance, because they can carry the inferential load internally. Yet as tasks become more complex (e.g., longer horizons, or greater combinatorial search), effective use depends on scaffolding, analogous to the role of Chain-of-Thought prompting in non-reasoning models (Wei et al., 2022; Zhou et al., 2023), where the human prompter has to perform part of the cognition by adding hints or heuristics that reduce the LLM's search space. This creates a new educational opportunity: students can learn when concise task descriptions suffice and when additional scaffolding is required.

However, providing hands-on, equitable practice at scale is difficult. Individual access to paid LLM tools, heterogeneous devices, institutional restrictions, and the logistics of running robotics hardware can prevent instructors from giving every student a comparable experience. What is needed are scalable learning activities, allowing entire classes to participate without needing a robot for every student, that (1) make model limitations observable, (2) teach prompt refinement across tasks and verification strategies, including how to distribute cognition effectively between humans and models, and (3) engage large numbers of students. To meet these requirements, we adopted principles from active learning (Bonwell & Eison, 1991; Freeman et al., 2014; Prince, 2004) and gamification (Deterding et al., 2011; Hamari et al., 2014), by designing an activity in which students tested and compared prompts while progress was reinforced through goals, scoring, and performance feedback.

This paper presents a scalable, gamified workshop designed for Master's-level robotics students to practice human-LLM collaboration in navigation planning. Students used their own smartphones to submit prompts through a lightweight web interface to two LLM planners: a non-reasoning model and a reasoning model. Their goal was to generate a robot-executable plan for grid-based navigation tasks of increasing difficulty. Concretely, students prompted the model to produce stepwise movement commands that would move a mobile robot through a grid while respecting constraints such as obstacle avoidance, visiting named checkpoints, and (in harder tasks) following a specified checkpoint order and avoiding revisiting cells. The interface returned an executable plan, a trajectory visualization, and an automated score based on rule violations. This setup made the differences in required scaffolding visible. For simpler tasks, a concise prompt often sufficed with the reasoning model; the non-reasoning model needed more explicit input; and as complexity increased, even the reasoning model benefited from human hints. Instructors used these outputs to facilitate discussion: students predicted whether prompts would succeed, compared high- and low-scoring prompts, and reflected on which kinds of scaffolding helped each model type. The workshop concluded with a live demonstration on a Boston Dynamics Spot robot to connect the simulation-based planning exercise to real robot execution and to surface practical issues.

This paper addresses two questions. First, we evaluated whether this workshop format is an effective and scalable way to teach students how to work with LLMs for robotics-relevant planning, specifically whether it yields high engagement and high self-reported learning about model differences, verification, and collaboration strategies. Second, we analyzed what students actually did when prompted with increasing task difficulty: how their prompting strategies changed as they learned to externalize and redistribute cognition. To do so, we analyzed the prompts students submitted during the workshop and responses to an anonymous post-task questionnaire measuring engagement and perceived learning, which we expected to be high given the interactive, feedback-rich format.

In summary, we present a practical workshop design for teaching critical LLM use in engineering, report empirical data on engagement and learning, and analyze how students changed their prompting as tasks became more complex.

## Methods

### Participants

A total of 226 students participated in the workshop as part of an MSc course on Human-Robot Interaction at the Faculty of Mechanical Engineering at TU Delft. The students were enrolled in either the MSc Robotics program or the Biomechanical Design track within the MSc Mechanical Engineering program. Participants signed up for the workshop in 10 time slots. Each workshop had a maximum of 30 participants. The number of participants per workshop ranged from 9 to 29. The study was approved by the Human Research Ethics Committee of TU Delft (approval number 6189).

### Hardware and software

We set up an area for real robot navigation, presentation, and interaction (see Figure 1). Two TVs were used: one displayed the presentation slides and the other was connected to an instructor’s laptop to show interfaces for interaction with students. A 7×13 virtual grid and a corresponding physical grid were set up for the navigation tasks. This grid contained 18 obstacle cells and three or four checkpoint cells. The navigation task was simplified as movements from cell to cell, represented in integer coordinates (row, column).

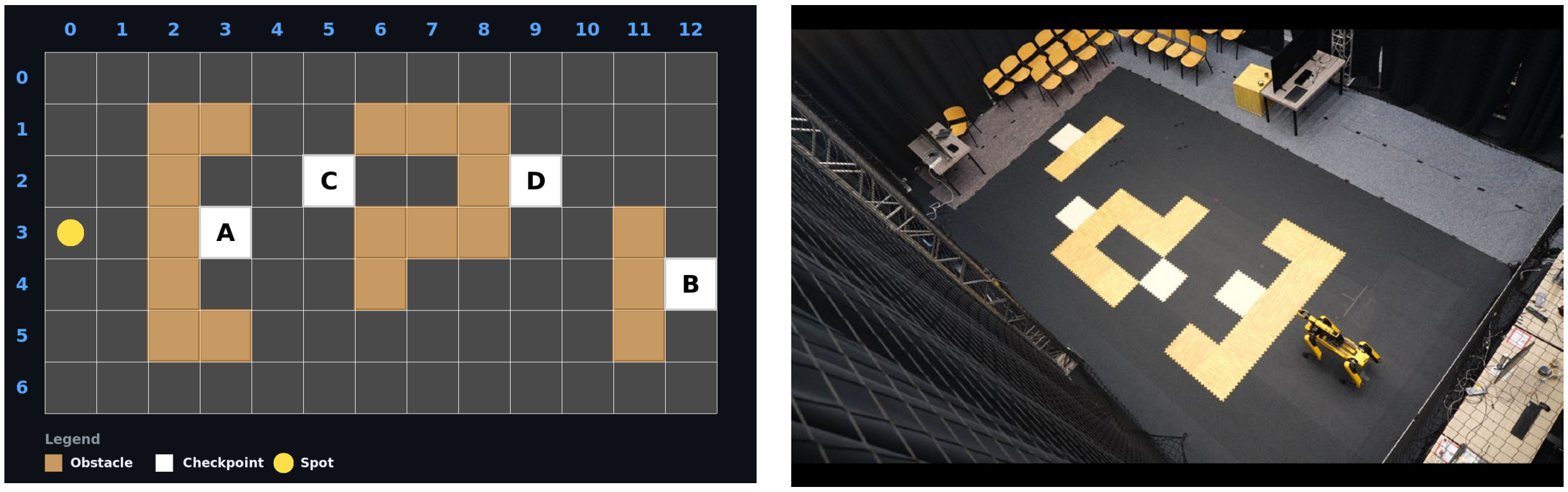

*Figure 1*. Left: Grid area for simulated robot navigation. Right: Corresponding grid area for real robot navigation.

The instructor's laptop (Dell Latitude 7320) ran a custom-built graphical user interface (GUI). The GUI was developed using Python's *Tkinter* GUI toolkit to visualize the simulated environment, run the LLM planners, display students' prompts, visualize planned trajectories, control the Spot robot in the demonstration, and show the position of the Spot robot in real-time (Figure 2, bottom). An instructor operated the GUI to process all user prompts collected from the students for each task in parallel and to show students' inputs and the LLM planner's output (Figure 2, top left).

User prompts were collected through a custom web interface hosted on the instructor's laptop. The host web interface tracked the progress, collected prompts, and displayed the final leaderboard (Figure 2). The client web interface on students' devices collected the students' prompts and showed the output plans from the LLM, together with an evaluation and visualization (Figure 3). Students accessed the client interface via a local wireless network provided by a TP-Link AX3000 Gigabit Wi-Fi 6 Router. Using the host interface, the instructor determined which client interface participants saw at each stage of the workshop.

The physical robot, Spot Explorer (Boston Dynamics, 2020), was controlled from the same laptop using the Python interface of Spot-SDK 5.0.1 (Boston Dynamics, 2025). The Noetic distribution of the Robot Operating System (Quigley et al., 2009) was used to bridge the GUI and the Spot SDK, enabling execution of the planned movements on the physical robot. Each LLM-planned move in the format "*move_to_cell row=<int> col=<int>*" was translated from the coordinates of a cell in the grid to a real-world target location in the world frame.

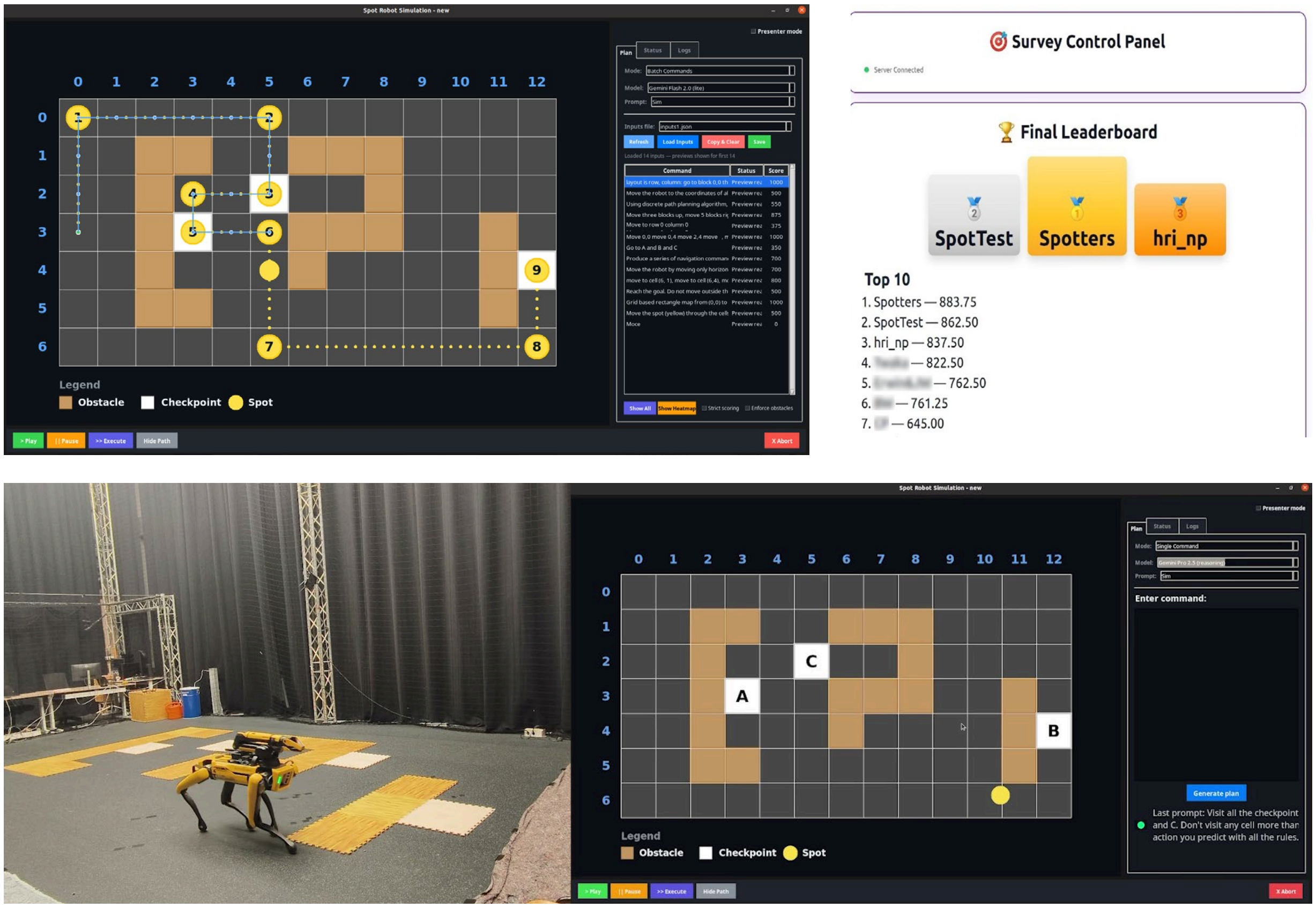

*Figure 2.* Top left: The simulation GUI displaying a trajectory generated by the LLM planner. Top right: The host web interface showing a leaderboard of mean student teams' scores. Bottom: The physical experimental setup, where the GUI (visible on the TV screen) provides real-time visualization and control of the robot during the demonstration.

A system prompt was combined with each user prompt (i.e., the prompt that students created during the workshop) to formulate a request to the LLM API server. The system prompt consisted of (1) the LLM's role: the LLM should act as a path planner on a 2D grid; (2) constraints: the planner should follow rules such as not going through obstacles and only moving horizontally or vertically on the grid; (3) the required output format: the planner had to output a robot-executable plan in the format "*move_to_cell row=<int> col=<int>*" per line; (4) a world description: a description of the grid dimensions, the robot's start coordinate, the checkpoints with names and coordinates, the obstacle coordinates, and the orientation of the grid; and (5) a guardrail: for impossible or non-relevant requests, the planner would not make a plan and respond "*FAIL <REASON>*" (see the full system prompt in Appendix A).

Two LLMs were chosen as planners: a non-reasoning model, Gemini 2.0 Flash-Lite (Google, 2025a), and a reasoning model, Gemini 2.5 Pro (Google, 2025b). All API requests were sent with a temperature of 0.0. For requests with Gemini 2.5 Pro (Google, 2025b), a 'thinking budget' of 3000 tokens was set to ensure an acceptable inference time during the workshop. This acted as a soft upper limit, meaning actual token usage varied based on the specific task (Google, 2025d).

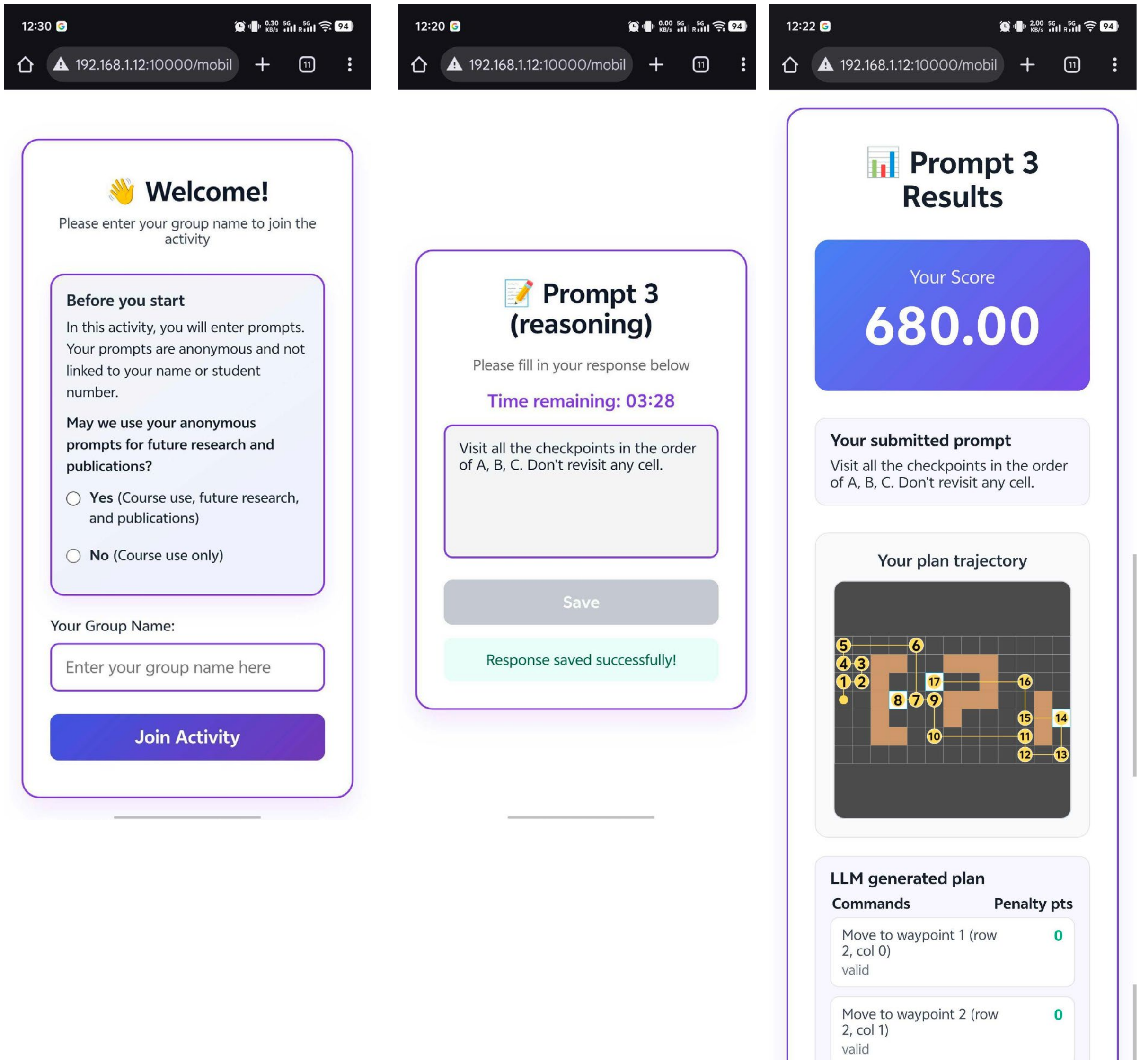

*Figure 3.* Screenshots from the client web interface, which students could access through their phones. Left: Welcome and consent screen; Middle: Prompt submission window; Right: Prompt results.

**Tasks**

We designed three tasks for students to learn by prompting the LLM planners on their phones via the web interface. The tasks used the same simulated environment but varied in difficulty.

- Task 1 was the simplest task: there were three checkpoints (A, B, C) to visit in any order. Students were asked to submit one prompt for the non-reasoning model (Task 1-NR) and one prompt for the reasoning model (Task 1-R).
- Task 2 was more difficult because there were additional constraints: the mobile robot had to visit checkpoints A, B, and C in that specific order, without revisiting any cell. Students were asked to submit a prompt for the reasoning model (Task 2-R).

- Task 3 was the most difficult task; it was based on Task 2 but involved one additional checkpoint (D) to visit. Again, students were asked to submit a prompt for the reasoning model (Task 3-R).

Each plan started with a baseline score of 1000, from which penalties were deducted whenever an action violated a constraint such as colliding with an obstacle or moving illegally. The GUI applied these penalties to the LLM-generated plans. Tasks were delivered via presentation slides (e.g., see Figure 4).

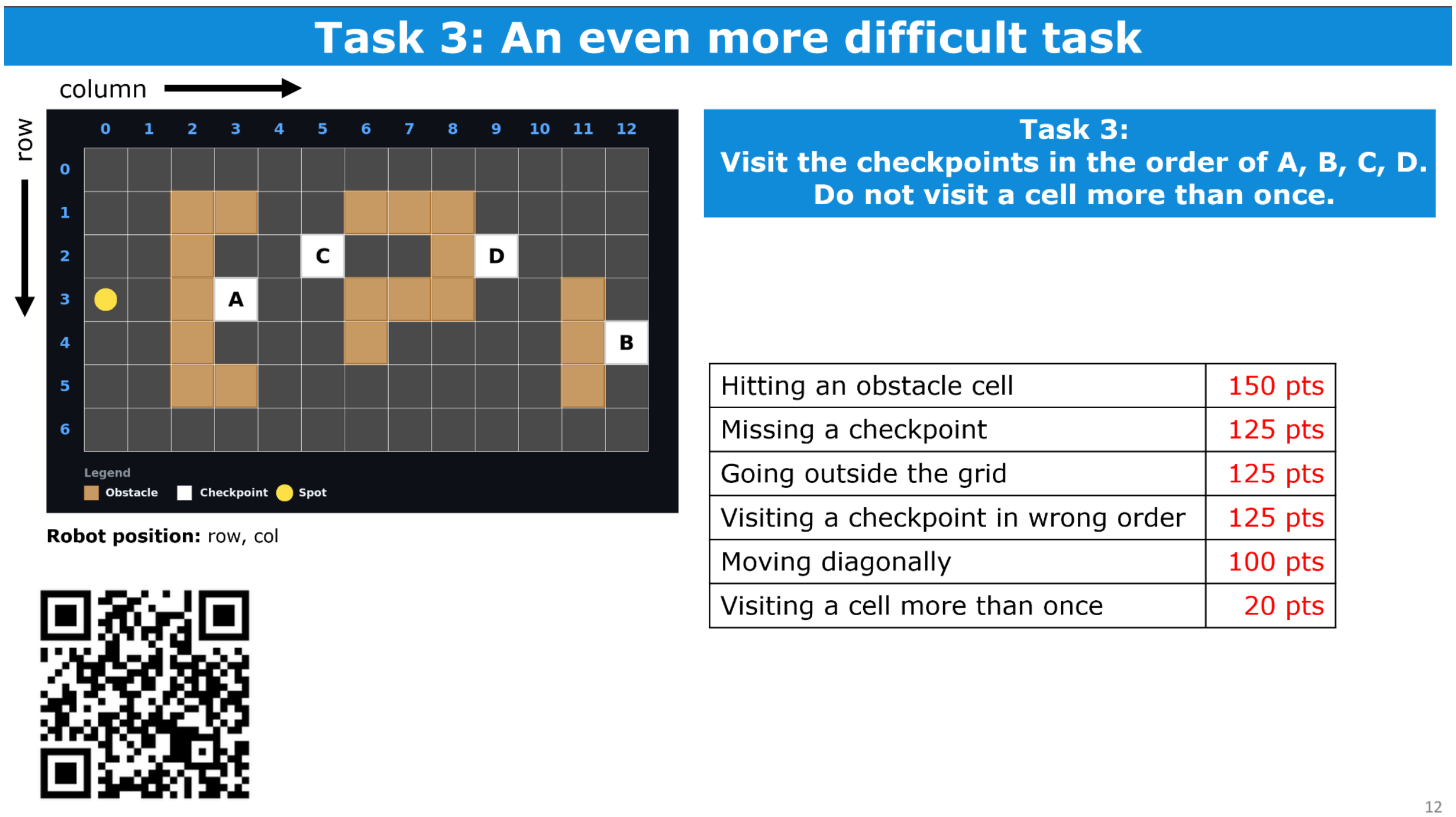

*Figure 4.* Presentation slide used to present Task 3.

We validated both the task difficulty and the importance of human hints beforehand using simulations. These simulations evaluated the baseline performance (a prompt consisting of the task description only) of the models. The non-reasoning model in Task 1 consistently generated plans with penalties, while the reasoning model was often able to solve Task 1 perfectly without human guidance. However, the performance of the reasoning model declined in Tasks 2 and 3 as the difficulty increased. Figure 5 presents an example plan from the reasoning model with a basic task description in Task 3.

We hypothesized that providing guidance on how to reach a checkpoint would improve the reasoning model's performance in Tasks 2 and 3, as such guidance reduces the planner's search space. To verify this, we evaluated Task 3 using prompts with varying levels of guidance, ranging from directions for the first checkpoint to directions for all four. The results supported our hypothesis: more guidance generally led to better performance. For example, a prompt containing "*... Go to A from the top and to B from the bottom.*" achieved a mean score of 650 (*SD* = 403; *n* = 10 repetitions; mean tokens used = 6,914), compared to the baseline mean score of 355 (*SD* = 395; *n* = 10 repetitions; mean tokens used = 13,471).

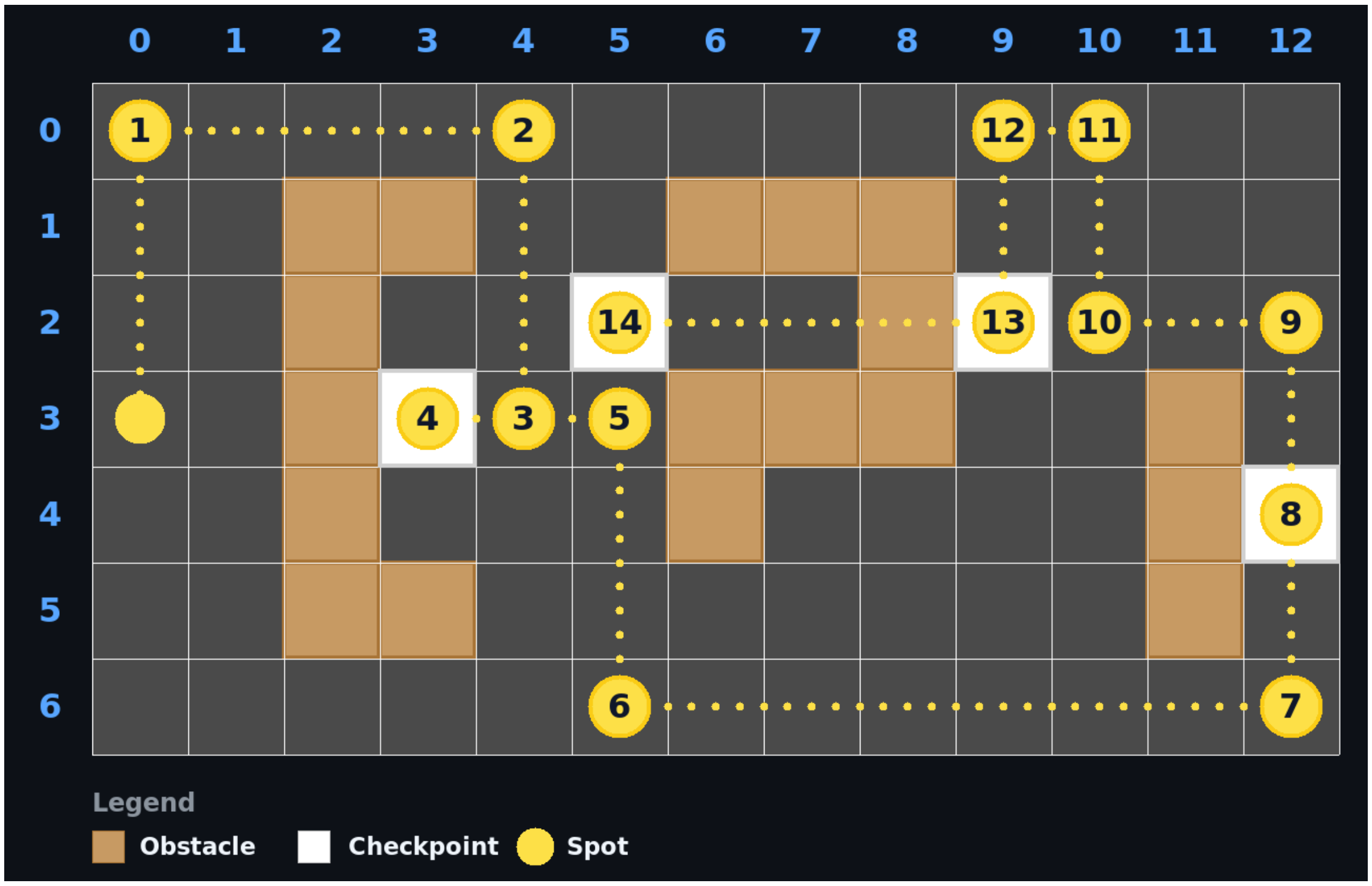

*Figure 5.* GUI screenshot visualizing a Task 3 trajectory plan produced by the reasoning model from a basic task description only. Two types of penalty occurred: colliding with an obstacle (from “13” to “14”) and revisiting two cells (from “4” to “5”). The score of this plan was 810 (1000 − 150 − 2 × 20).

To highlight differences in model capabilities across tasks when interacting with students, we designed one example prompt per task condition. In Task 1, we provided the non-reasoning model with coordinate-based, step-by-step instructions (e.g., “*Go to (0,0), then go to (0,4), then go to (3,4) …*”), while the reasoning model received only a basic task description. In Task 2, we used a basic task description augmented with a “self-check” instruction (e.g., “*…Check every action you predict with all the known rules*”)[1]. In Task 3, we supplied a basic task description with directional hints (e.g., “*…Approach A from the top and B from the bottom*”).

**Workshop flow**

The workshop was designed as an 80-minute session. The protocol evolved slightly after the first two workshops based on feedback about login timing, when and how to present the system prompt, and how to run the robot demonstration. It was also noted from the first two that working in teams was more engaging than working individually. The workshop was led by two instructors:

[1]It should be noted that while self-correction methods can improve accuracy for standard models (Tie et al., 2025), their reliability varies across tasks; standard models frequently show even degraded performance under intrinsic self-correction (Huang et al., 2024). When applied to modern reasoning LLMs, these interventions yield only marginal gains relative to their computational cost, as reasoning models natively execute internal error-detection and correction steps during generation (Tie et al., 2025). In summary, the fact that this specific ‘self-check’ prompt worked in the reasoning context of Task 2 should not be taken as evidence that such prompts are a generally recommended strategy.

one delivered the presentation and the other managed the setup, while both facilitated the discussion on prompts.

Students were first introduced to the learning goals, specifically focusing on the background of LLMs and the principles of controlling robots via natural language. Next, the students joined a local Wi-Fi network by scanning a QR code to access the web interface on their phones, mostly in teams of two. After scanning a second QR code to enter the interface, students provided an alias and indicated whether they consented to the use of their data for research purposes.

Then, an interactive phase took place, consisting of three tasks where the students created prompts. The instructor explained the system prompt and introduced Task 1 by explaining the task goal, the gameplay (an initial score, penalties, and the leaderboard), and the LLMs used. Students were given eight minutes to draft their two prompts for Task 1. Prompts from all student teams were then collected via the web interface and sent to the LLM API in parallel via the GUI.

While complete plan generation of all submitted prompts for the non-reasoning model required only a few seconds, the reasoning model typically took approximately 40 seconds, and occasionally several minutes. When waiting for plans to be generated, the instructor facilitated a discussion with students using several student-written prompts, encouraging them to predict the performance. Once plan generation and evaluation were complete, the instructors led a further discussion using visualizations of the generated plans to analyze why certain prompts worked better than others. The cycle ended with the instructor showing one of the four example prompt outputs, after which students reviewed their prompt outputs, consisting of the text-based plan, visual trajectories, scores, and detailed penalty breakdowns, on their phones and discussed with instructors if they wanted to. The total duration of the discussions was approximately 12 minutes (6 minutes for Task 1-NR; 6 minutes for Task 1-R).

The cycle of briefing, prompt drafting, and discussion was repeated for Tasks 2 and 3, with a prompt-drafting time limit of 4 minutes and a mean discussion time of about 8 minutes for each of the tasks. After completing the interaction of all three tasks, the instructor showed a final leaderboard, depicting student teams' mean scores across the four generated prompts, via the host web interface on the TV screen (Figure 2).

Following the tasks, the instructor summarized key takeaways regarding human guidance in planning tasks and provided transparency on the computational costs per prompt. The instructor explained that for Task 1, compared to the non-reasoning LLM, the reasoning LLM used many more tokens and therefore incurred higher expenses. On the upside, it was emphasized that the reasoning model allowed for high-level communication and offloading of the path planning to the LLM.

Students were then invited to complete an anonymous feedback questionnaire consisting of four questions. The workshop culminated in a 10-minute live demonstration of a Boston Dynamics Spot robot. The robot executed a real-world, penalty-free trajectory based on a student-generated plan from Task 2. The demonstration included manipulation: the robot was commanded to "*grasp* *the tomato can in front of you*" using Gemini 2.5 Pro (Google, 2025b), demonstrating that LLMs can handle multimodal input, i.e., text and image data. The instructors explained the reality gap using robot localization as an example, mentioning the sensor offset, robot locomotion, and the transformation from the grid frame to the real world, and how LLMs can also be used for perceptual tasks.

**Data collection and analysis**

We collected all prompts submitted by students via the web interface, including the LLM-predicted plans and token usage. The evaluations of the students' prompts were analyzed using three measures: (1) the penalty-free rate, defined as the proportion of prompts resulting in LLM plans that had no penalty events; (2) the no-plan rate, defined as the proportion of prompts for which the LLM output "FAIL"; and (3) penalty counts, defined as the number of penalties triggered within a plan (see Figure 4). No-plan prompts were excluded when reporting the penalty counts.

The feedback questionnaire included two Likert-scale questions and two open-ended questions:

Q1: *To what extent did the workshop help you understand the differences between non-reasoning LLMs and reasoning LLMs?* (A great deal, A lot, Moderately, A little, Not at all, Prefer not to say)
Q2: *How engaged did you feel during the workshop?* (A great deal, A lot, Moderately, A little, Not at all, Prefer not to say)
Q3: *What were your main strategies for formulating prompts so that the LLM would understand your intent?*
Q4: *Do you have any other feedback on the workshop? What was the most valuable part, and what changes could we make to improve it?*

The students' prompts and answers to Q3 and Q4 were analyzed using LLM-assisted quantitative content analysis by Gemini 3 Pro (Google, 2025c). Each prompt/answer was evaluated against pre-defined categories using a binary (0 or 1) classification, allowing items to fall into multiple categories or no category at all. The final codebook for the students' prompts, Q3, and Q4 contained 11, 10, and 15 categories, respectively. The full codebooks are provided in Appendix B.

## Results

**Prompts generated by students**

Across all workshops, a total of 494 prompts (*n* = 123, 123, 124, 124 for Task 1-NR, Task 1-R, Task 2-R, Task 3-R, respectively) were collected from students who provided consent. Not all collected prompts were evaluated by the LLM due to the API rate limit and rare non-responses. Therefore, the performance analysis was conducted on 457 prompts (*n* = 123, 122, 101, 111) with valid evaluation results.

Figure 6 summarizes the performance of the students' prompts across the four conditions. For Task 1, the non-reasoning model achieved a penalty-free rate of 18.7%, whereas the reasoning model achieved 55.7%. This performance improvement is also reflected in the penalty counts (median of 3 versus 0). Performance decreased as task difficulty increased from Task 1 to 3. Specifically, the penalty-free rate was 30.7% for Task 2 and 21.6% for Task 3, with a median number of penalties of 1 for both tasks. Furthermore, Task 3 also showed the highest no-plan rate of 19.8%. This was primarily caused by the model's inability to find a solution during its search, rather than the presence of irrelevant user prompts.

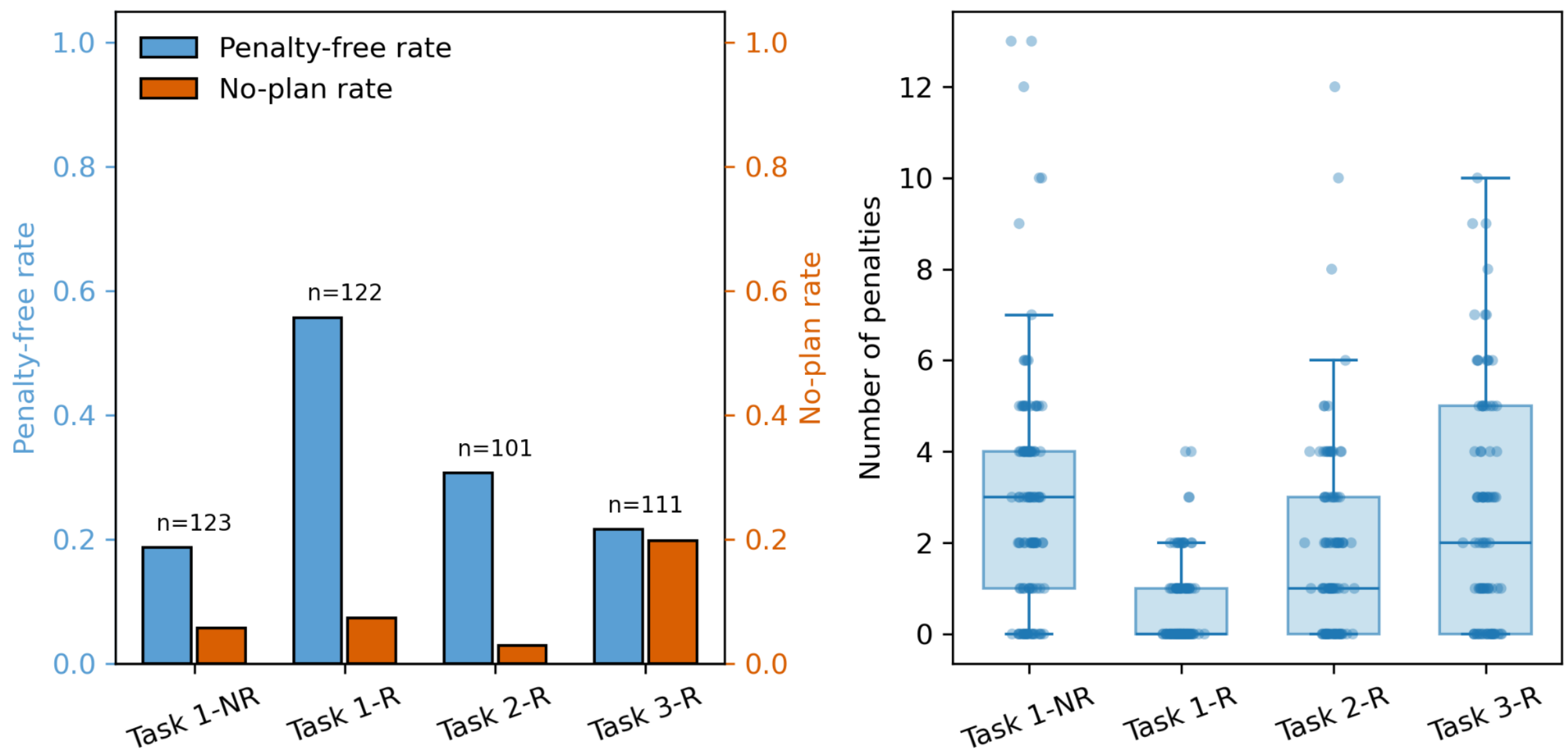

*Figure 6.* Performance of the students' prompts across tasks. Left: Penalty-free rate and no-plan rate for each prompt condition, with the number of evaluated prompts. Right: The distribution of penalty counts, shown as scatter points, for each prompt condition. NR: Non-reasoning LLM; R: Reasoning LLM.

We analyzed the mean frequency of each penalty type within each prompt condition (see Figure 7). Collisions were the dominant penalty type across all tasks. Compared with the reasoning model's performance in Task 1, the non-reasoning model triggered more collisions (1.681 vs. 0.389), more diagonal moves (0.612 vs. 0.035), and more out-of-bounds penalties (0.345 vs. 0.000). For the more complex tasks (Tasks 2 and 3), the second most common penalty was revisiting cells; the mean penalty count for revisiting increased from 0.684 in Task 2 to 1.213 in Task 3.

We also compared computational resource usage (tokens) and time across all tasks (see Figure 8). Requests and responses in Task 1-NR were fast and low-cost (median: 1,032 tokens; 1.2 s latency). In contrast, reasoning prompts consumed substantially more resources (Task 1-R median: 5,584 tokens; 38.1 s). Consumption rose as task difficulty increased (Task 3 median: 7,167 tokens; 51.6 s), driven by the model's greater use of "thinking tokens" (see Appendix C).

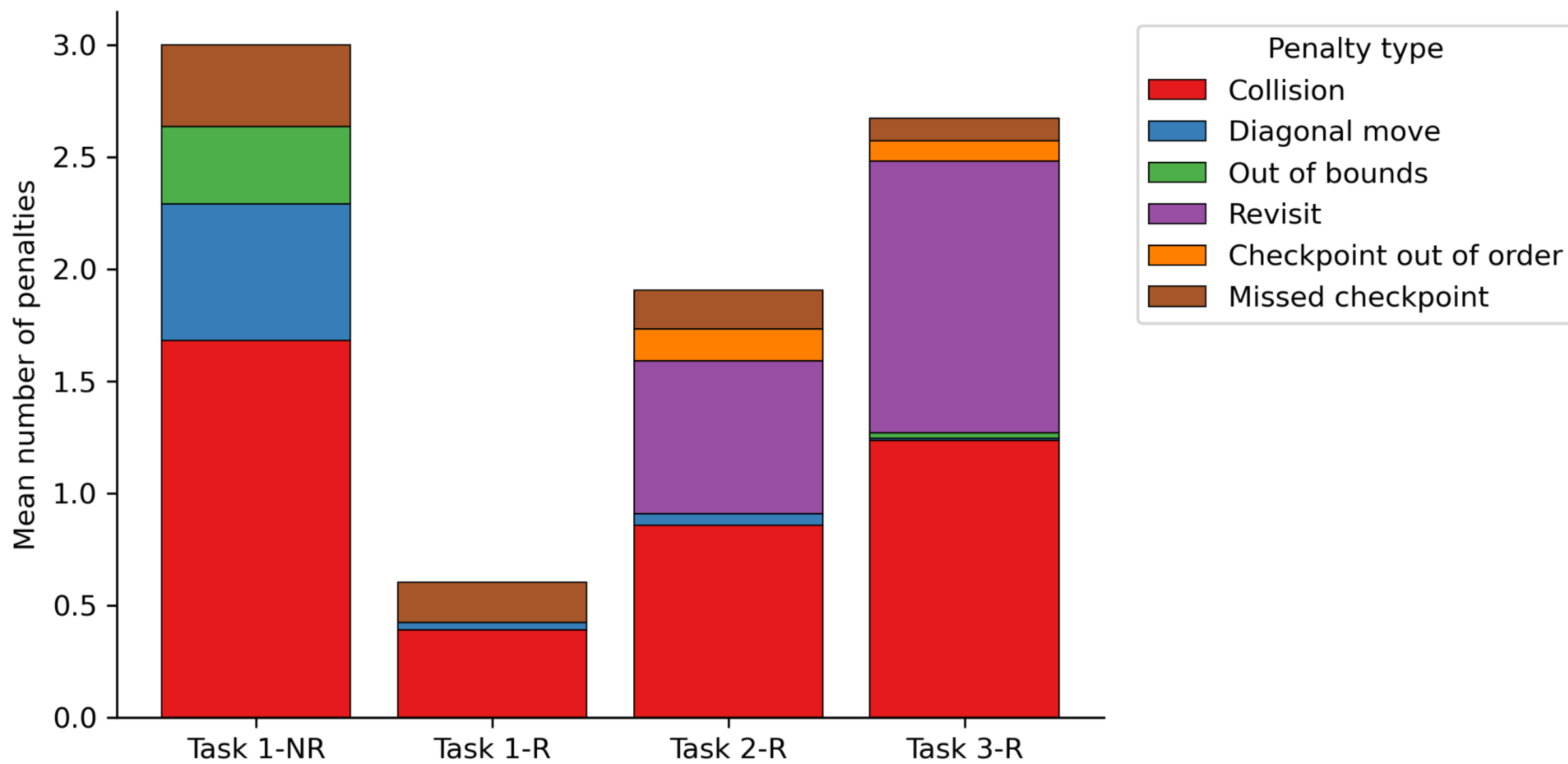

*Figure 7.* Mean number of penalties per plan, categorized by penalty type. NR: Non-reasoning LLM; R: Reasoning LLM.

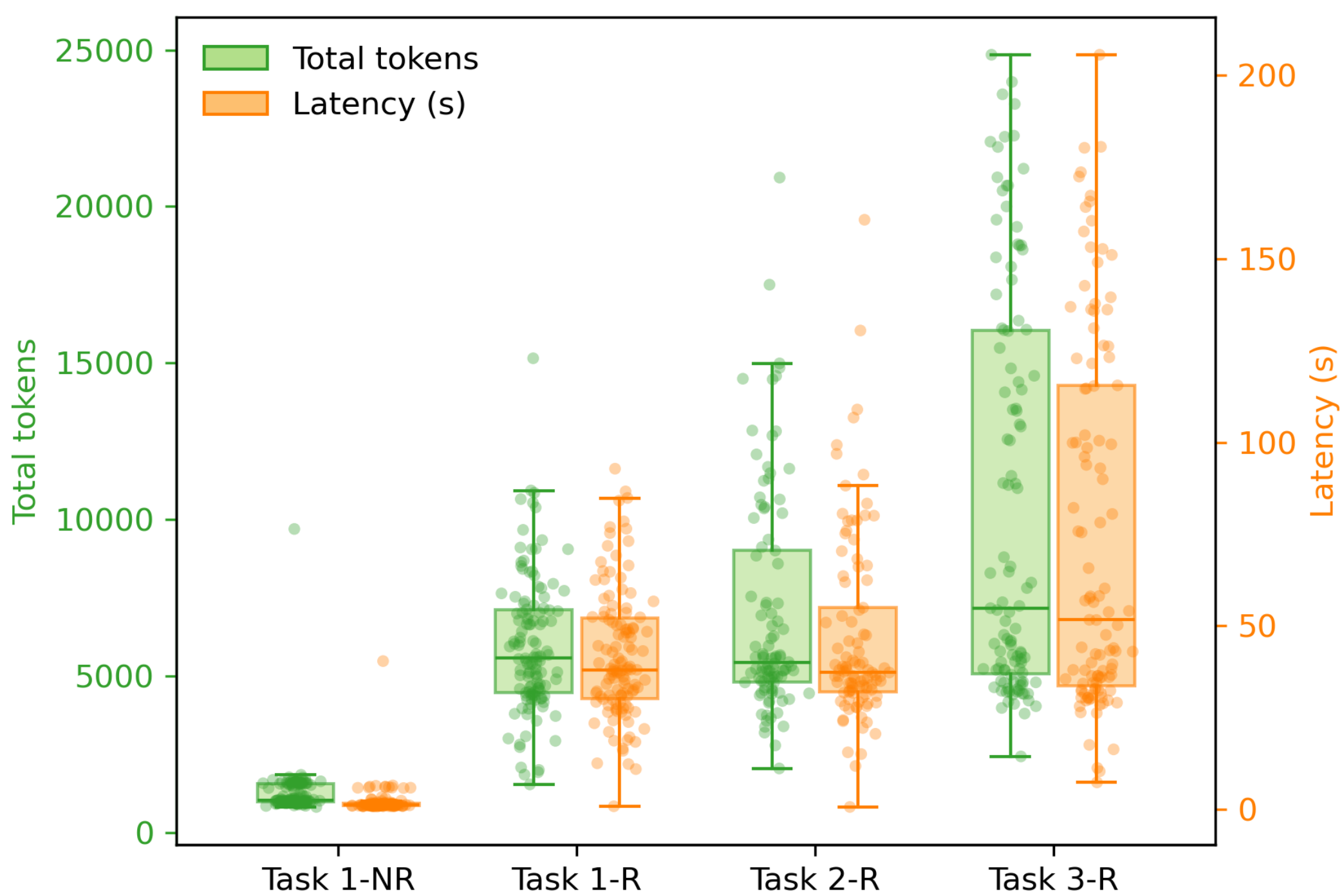

*Figure 8.* Token usage and latency during the workshop. NR: Non-reasoning LLM; R: Reasoning LLM.

We further analyzed the linguistic patterns of the student-generated prompts, focusing on how instructions, constraints, and human guidance were provided within the conditions. Prompting patterns varied by prompt condition (see Figure 9). In Task 1, prompts for the non-reasoning model frequently micromanaged navigation with explicit human-planned paths (48.8%), e.g.,

"*…First move in the straight line to (0,0), Then turn right and move 4 cells*...", while prompts for the reasoning model contained fewer such step-by-step instructions (18.7%). As task difficulty increased, students started to provide high-level guidance. In Task 3, explicit step-by-step routes were rare (3.2%) and students provided more partial waypoints (37.1%), e.g., "*First go to 6,0 and then to 6,4. Then go to A and after to C…*" or strategic heuristics (23.4%), e.g., "*...Hint: the path consist [sic] 2 counterclockwise spirals*…" They also often asked the model to self-check before outputting a final plan (40.3%), e.g., "*Validate each move*".

The changes in prompting strategies from Task 1 to Task 3 were consistent with the key messages and example prompts the instructor provided after each task: After Task 1, the instructor explained that the non-reasoning model required step-by-step guidance whereas the reasoning model could perform well with only concise user input of the task description. Such information may have contributed to the decrease in "*Constraint repetition*" and "*Path explicit step-by-step*". After Task 2, the instructor explained the idea of providing hints or self-checks for complex tasks, which may have resulted in increased usage of "*Path partial waypoints*", "*Path strategic heuristic*", and "*Reasoning self-correction*" in Task 3-R.

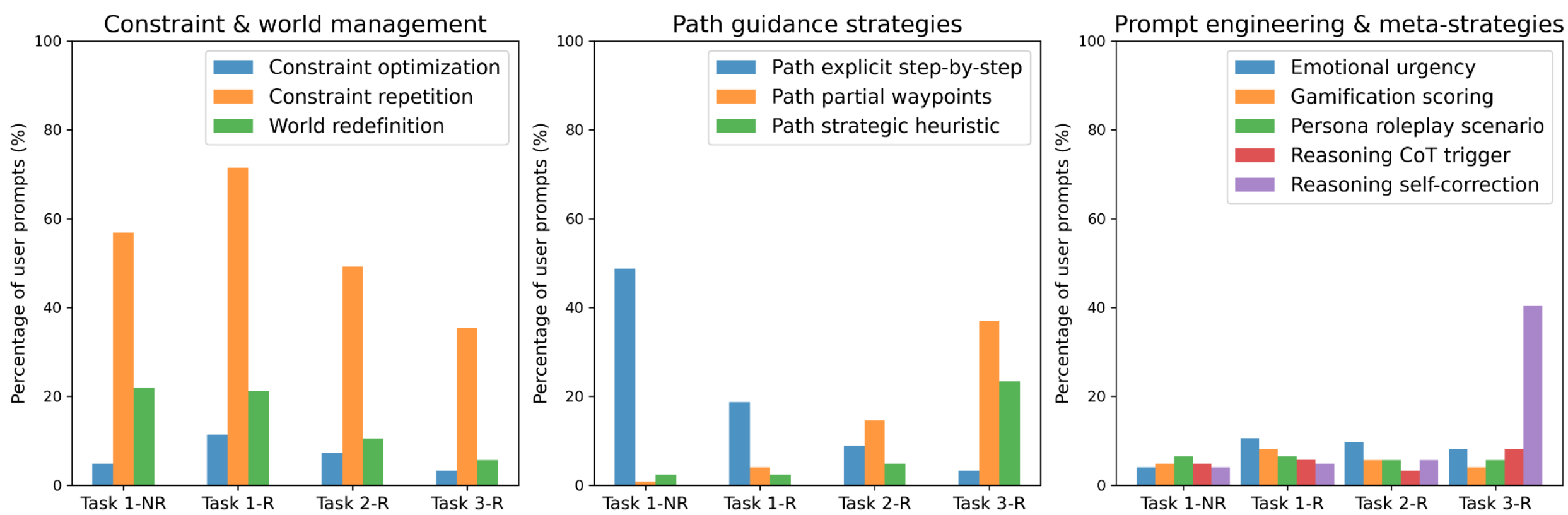

*Figure 9.* Distribution of the categorization of students' prompts, by task. NR: Non-reasoning LLM; R: Reasoning LLM.

In an exploratory analysis, we examined the rank correlations (Spearman's ρ) between the coding of students' prompts and penalty counts incurred. For this analysis, no-plan prompts were assigned a high penalty count of 10 to reflect their failure to generate a plan. The correlations between the binary coding results (4 task conditions × 11 categories) and penalty counts revealed statistically significant associations for certain prompt types. Providing step-by-step guidance with the non-reasoning model in Task 1, such as "*Go 3 steps down Go 12 steps right Go 2 steps up…*", was associated with fewer penalties ($\rho = -0.437$, $p < 0.001$, $n = 123$), as were heuristic strategies when using the reasoning model in Task 3 (e.g., *"Go to A through row zero, then move in a spiral way to B through row 6*"; $\rho = -0.240$, $p = 0.011$, $n = 111$). In contrast, prompts containing additional (and unneeded) optimization constraints (e.g., "*...Find path with minimal amount of state changes…*") with the reasoning model in Task 2 resulted in higher penalty counts ($\rho = 0.196$, $p = 0.049$, $n = 101$). Repeating known constraints had mixed effects: a positive correlation for the non-reasoning model in Task 1 ($\rho = 0.210$, $p = 0.019$, $n = 123$), but negative correlations with the reasoning model in Tasks 1 and 2 (ρ = -0.213, -0.203; $p$ = 0.018, 0.041; n = 122, 101). All other correlations were not significant ($p > 0.05$).

**Post-task questionnaire**

In total, 214 questionnaire responses were collected. We included in the research analysis only responses from participants who consented to the use of their data for future research and publication ($n$ = 200). The response distributions of Q1 and Q2 are visualized in Figure 10. A majority reported high perceived learning and engagement: 81.5% selected "A lot" or "A great deal" for understanding model differences, and 91.0% selected "A lot" or "A great deal" for engagement (Figure 10).

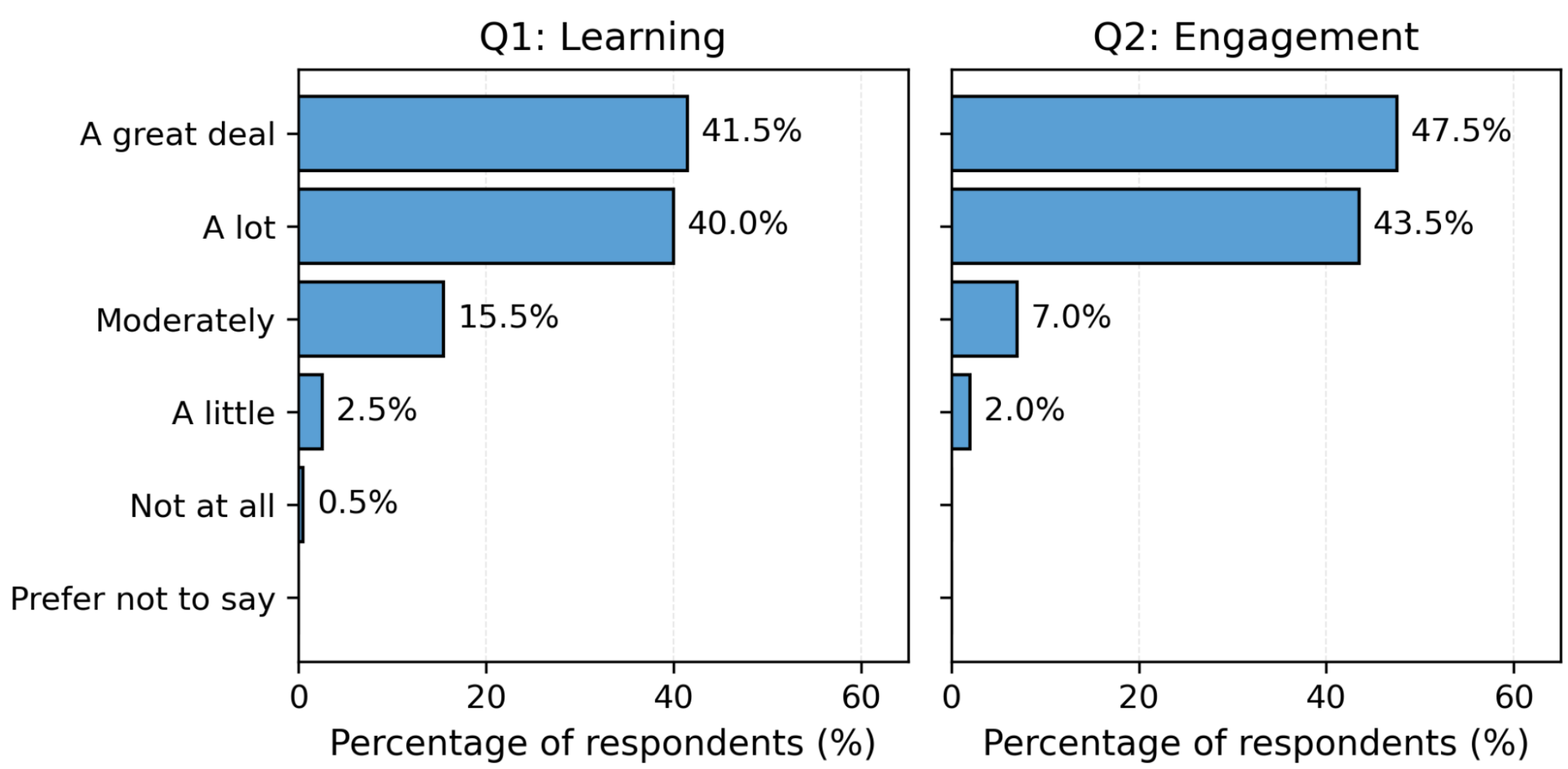

*Figure 10.* Response distributions for Question 1 (Q1, left) and Question 2 (Q2, right) of the post-task questionnaire ($n$ = 200).

For the open-ended question Q3 ("*What were your main strategies for formulating prompts so that the LLM would understand your intent?*"), we excluded six empty submissions, leaving 194 responses for analysis. Among 10 final coded strategies, the most frequently reported strategies were using explicit, unambiguous task descriptions (33.5%; e.g., "*be as precise as possible*") and keeping prompts clear and concise (29.9%; e.g., "*Be clear and concise*"). Beyond these, reiterating rules and key instructions was also common (19.1%; e.g., "*Repeating rules & instructions*"). Participants further described two levels of guidance in their strategies: providing partial plans or hints (18.6%; e.g., "*Give direction to follow*") and, less often, providing complete solution steps (14.4%; e.g., "*instruct the LLM to follow the path*").

Students were also asked to respond to the open-ended question: "*Do you have any other feedback on the workshop? What was the most valuable part, and what changes could we make to improve it?*" (Q4). A total of 137 responses were provided to this question. Responses were mostly positive, consistent with Q1 and Q2. A total of 43.1% used the text field to express general satisfaction (e.g., "*I liked it!*"), while 10.2% explicitly stated they had no additional feedback or critique to offer (Figure 11).

When asked about the most valuable parts, participants mostly highlighted the pedagogical value over the hardware. The visualization of results was the most frequently mentioned feature (13.1%), with participants appreciating "*seeing how [their] prompts performed*" in the simulation. This was followed by the comparative analysis of prompts (11.7%), with students noting that "*the discussion about why each prompt failed and worked proved to be valuable.*" In addition, 6.6% of participants recognized the value of hands-on experience in prompting, commenting that the "*Most valuable [aspect] was to do it ourselves*".

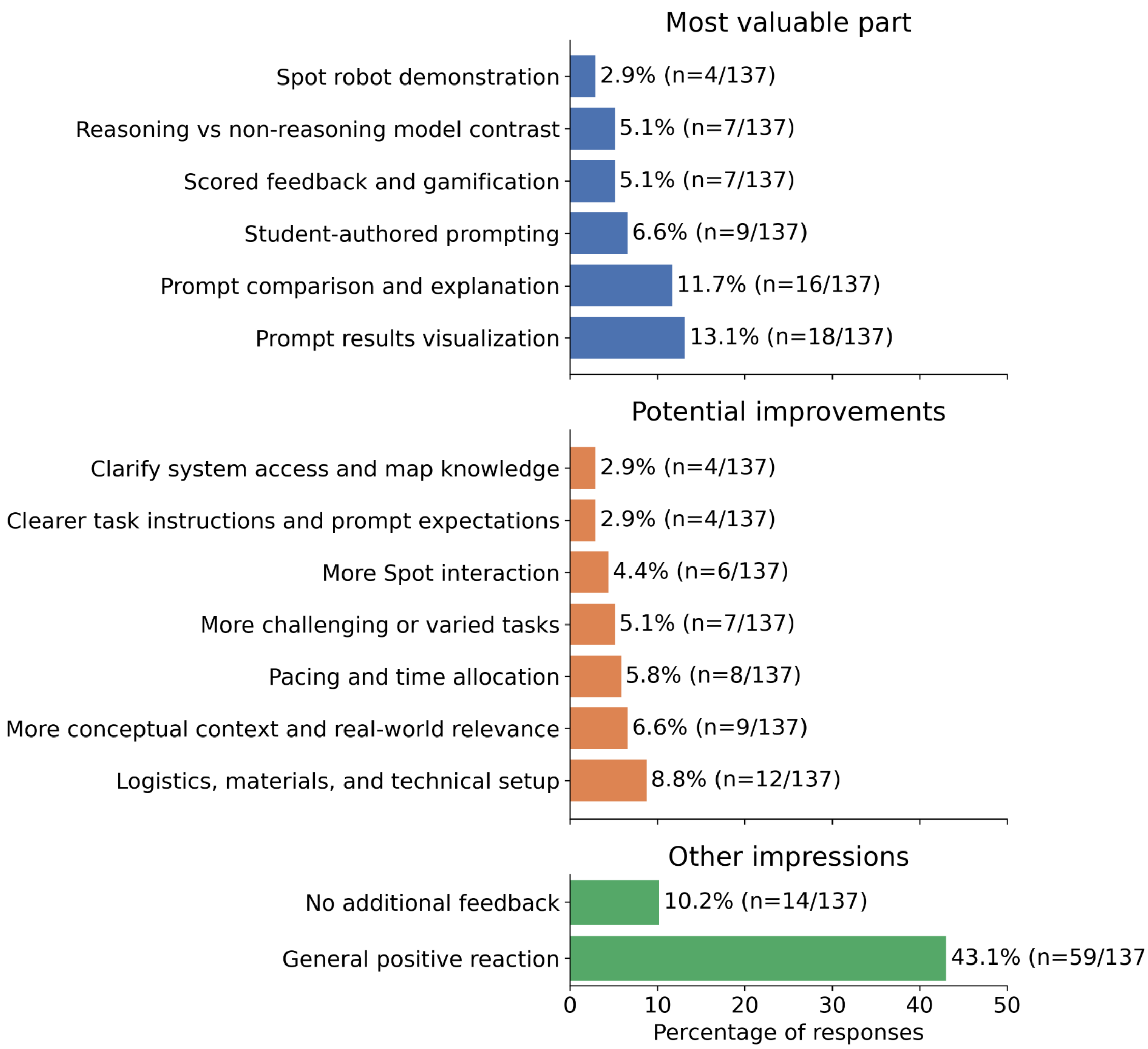

*Figure 11.* Distribution of the students' responses to Question 4 ("*Do you have any other feedback on the workshop? What was the most valuable part, and what changes could we make to improve it?*"; Q4) of the post-task questionnaire.

Suggestions for improvement clustered around logistical friction and conceptual depth. The most frequent issue (8.8%) concerned the setup of the workshop, mentioning the sitting space ("*Create wider places to sit* :)"), view of the screen ("*...a larger screen...*"), and technical issues such as the API rate limit and the QR-code-based login. Pedagogically, 6.6% of students requested greater real-world context, asking for examples of "*how LLM based robotics is used... to perform real tasks*" to better understand the practical utility of the exercises. Because the demo occurred after the questionnaire, students were not yet aware of the real-world applications being presented. Additionally, some students felt the pacing and difficulty could be adjusted; while some noted the "*time between the tasks is quite [long]*", others suggested the exercises "*could be even more difficult/challenging*" for the LLMs as the initial tasks were "*quite simple*".

## Discussion

This study set out to answer two questions: (1) whether a scalable gamified robotics workshop can teach critical LLM use at cohort scale, and (2) how students' prompting strategies change as task constraints increase. Robotics is well suited for this because an LLM's output can be directly executed, and any errors show up as observable robot behavior.

By placing students in a gamified setting, we found that hands-on exposure to model limitations was associated with high self-reported learning about verification and model differences. The workshop's success was revealed by high levels of student engagement and perceived learning; questionnaire responses indicated that 91.0% felt highly engaged, while 81.5% reported learning "a lot" or "a great deal" about the distinctions between model types. This was achieved with a lightweight, bring-your-own-device setup, making the format easy to replicate in other courses. In total, processing the 494 prompts cost an estimated 23.19 USD, and median end-to-end latency ranged from 1.2 s (Task 1 non-reasoning) to 51.6 s (Task 3 reasoning), making cost and delay concrete constraints that students had to account for.

The prompt data show that students shifted their strategies in a way consistent with a distributed-cognition view of human-AI collaboration. As task complexity increased relative to model capability, students came to redistribute the inferential load. They moved from step-by-step guidance for non-reasoning models toward higher-level guidance when using a reasoning model in a complex task, i.e., using waypoints or strategic heuristics to reduce the model's search space. Students appeared to calibrate their strategies based on the instructor's guidance and their own experience, learning when a brief task description was enough and when they needed to provide more structure.

From a robotics perspective, while the workshop used a direct "LLM-as-planner" architecture (Chen et al., 2024; Huang et al., 2022; Silver et al., 2024; Zhou et al., 2024) to deliberately expose students to the consequences of unverified outputs, such as collisions, it is important to acknowledge that responsible professional deployment requires a safer architecture. In real-world applications, relying solely on the probabilistic output of an LLM for motor control would be irresponsible at the current state of technology. As discussed during the workshop, a viable alternative to the direct prompting used in our exercises is "LLM-informed classical planning" (Meng et al., 2024; Xiao & Yamasaki, 2025), where the LLM is used for its semantic understanding and high-level logic, while the actual trajectory generation is delegated to rule-based solvers. In addition, an algorithmic safety layer could ensure that, regardless of the LLM's hallucinations, the robot acts within physically feasible bounds, preventing collisions (Ravichandran et al., 2026).

Several limitations should be noted. The use of a discrete grid world abstracts away continuous dynamics in physical robotics, which simplifies the verification challenges students face. Additionally, the efficacy of the prompt strategies observed may depend on the specific model versions used (Gemini 2.5 Pro; Google, 2025b) and may evolve rapidly as reasoning capabilities improve. Furthermore, while a simulation-to-reality component was present via the Spot robot demonstration, logistical constraints meant that most student interaction occurred in simulation, limiting exposure to hardware capabilities such as battery limits, sensing uncertainty, and communication latency.

In conclusion, this work showed that scalable gamified environments can connect theoretical AI knowledge with practical engineering skills. By engaging students in hypothesis generation, testing, and correction, the workshop provides a practical model for teaching the distributed cognition required in modern human-AI collaboration. The role of the engineer will increasingly resemble that of a verifier and supervisor; curricula should therefore continue to emphasize that while AI can accelerate planning, the final check before physical action remains a human obligation.

## Appendix

### A. System prompt

In the workshop, a system prompt was sent together with every student-written prompt to the LLM API server. Below is the system prompt appended to the students' prompts.

```
### ROLE
You are a path planner for a mobile robot operating on a 2D grid. Your goal is to
generate a valid sequence of actions according to the USER_PROMPT.

### AVAILABLE ACTION:
- move_to_cell row=<int> col=<int>

### CONTEXT:
CURRENT_STATE = {current_state_json}
WORLD_MODEL = {world_model_json}

### RULES:
1. Start state:
- The robot begins at CURRENT_STATE.robot_cell.
- The left top cell is at (row=0,col=0) and the right bottom cell is at
(row=6,col=12).
2. Movement logic:
- The robot performs each move_to_cell by traveling in a straight horizontal or
vertical line.
- The robot should not navigate through any obstacle cell (listed in
WORLD_MODEL.obstacle).
- The robot should not move out of the grid.
3. Error handling:
- This system prompt is the ground truth. If a user's prompt is not relevant to the
role or suggests an action that is impossible (e.g., dance or backflip), output:
"FAIL <REASON>".

### OUTPUT FORMAT:
- Return only the action commands, one per line, or a FAIL message.
- Do not emit JSON, comments, numbering, braces, brackets, or quotation marks.

### EXAMPLES
User: "Go to the left top corner"
CURRENT_STATE.robot_cell: row=3 col=0
Ouput:
move_to_cell row=0 col=0

User: "Hello world"
Ouput:
FAIL Not relevant to the role.

### USER PROMPT
{command}
```

## B. Codebook generation and encoding pipeline

We conducted a quantitative content analysis of student prompts using a multi-stage coding process. To develop the initial codebook, we used Gemini 3 Pro (Google, 2025c) to analyze a random sample of 100 prompts (25 per condition). We refined this codebook iteratively by prompting the LLM with guiding questions on prompting patterns and validating the emerging themes, resulting in 11 final categories. These categories were then applied to the full dataset via binary multi-label coding. To mitigate potential LLM order effects, we randomized both the response and code sequences for each run.

The 11 prompt categories (Table B1) comprise three groups:

- Constraint & world management: Includes references to system constraints (Constraint repetition), user-specified parameters (Constraint optimization), and the redefinition of environmental coordinates (World redefinition).
- Path guidance strategies: Ranges from explicit navigational instructions (Path explicit step-by-step) to partial waypoints (Path partial waypoints) and high-level heuristics (Path strategic heuristic).
- Prompt engineering & meta-strategies: Encompasses chain-of-thought triggers (Reasoning CoT trigger), self-validation requirements (Reasoning self-correction), persona framing (Persona roleplay scenario), gamification (Gamification scoring), and emotionally charged language (Emotional urgency).

Following a similar iterative process, we developed codebooks for the open-ended responses in Q3 (10 categories; Table B2) and Q4 (15 categories; Table B3).

Table B1. *The codebook for the content analysis of students' prompts*

| **Category** | **Definition of category** |
|---|---|
| Path_Explicit_StepByStep | The user provides a complete (or nearly complete) sequence of specific coordinates, directional moves (up/down/left/right), or cell-by-cell instructions to solve the task, effectively treating the LLM as a transcriber rather than a planner. |
| Path_Strategic_Heuristic | The user provides a high-level geometric shape, pattern, or general navigation strategy to guide the pathfinding without providing a strict step-by-step coordinate list. |
| Path_Partial_Waypoints | The user suggests specific intermediate target cells, rows, or columns to visit between checkpoints to aid navigation, without providing the full trajectory. |
| Constraint_Repetition | The user explicitly restates rules that are already known to the system via the hidden system prompt (e.g., do not move diagonally, do not hit obstacles, stay within the grid). |
| Constraint_Optimization | The user imposes a requirement for efficiency or specific performance metrics that were not strictly required by the basic task description (e.g., shortest path, least steps, fewest turns). |
| World_Redefinition | The user explicitly describes the grid dimensions, obstacle locations, or coordinate system in the prompt, despite the system prompt already containing this data. |
| Reasoning_CoT_Trigger | The user explicitly instructs the model to think, reason, plan sequentially, or explain its logic (Chain of Thought) rather than just outputting the result. |
| Reasoning_Self_Correction | The user instructs the model to verify, double-check, or validate its generated path against the rules before outputting the final answer. |
| Persona_Roleplay_Scenario | The user assigns a specific persona to the AI (e.g., "Robot Planner") or creates a fictional narrative/scenario (e.g., "Prison escape") to frame the task. |
| Gamification_Scoring | The user introduces a specific point system, rewards, or penalties to influence the model's decision-making process. |
| Emotional_Urgency | The user employs capitalization, exclamation marks, pleading language, or emotionally charged words to emphasize the importance of following the rules. |

Table B2. *The codebook for the content analysis of Q3*

| Category | Definition of category |
|---|---|
| Conciseness & Simplicity | Strategies focused on keeping prompts short, direct, and free of unnecessary complexity to avoid confusing the model. |
| Specificity & Explicitness | Strategies involving highly detailed descriptions of the task environment, objectives, or boundaries to ensure no ambiguity. |
| Model-Specific Adaptation | Strategies that explicitly differentiate the prompting approach based on whether the model is a "reasoning" model or a "non-reasoning" model. |
| Constraint Emphasis & | Strategies that focus on reiterating rules, stating what *not* to do (negative constraints), or repeating key instructions to ensure adherence. |
| Step-by-Step Structuring | Strategies that break the task down into a sequence of operations, logical steps, or use code-like formatting to organize the prompt. |
| Low-Level Guidance | Strategies where the user performs the planning logic themselves and provides the LLM with specific coordinates, row/column numbers, or the exact path to follow. |
| High-Level Scaffolding | Strategies that provide clues, general directions, or partial plans to guide the model's reasoning without giving the full solution. |
| Anthropomorphism & Role- | Strategies that treat the LLM as a human (e.g., a child, a student), use emotional appeals, or create high-stakes fictional scenarios. |
| Self-Correction & Verification | Strategies that explicitly instruct the model to review its own work, double-check constraints, or reflect before finalizing the answer. |
| Iterative Learning | Strategies based on trial-and-error, adapting prompts based on previous failures, or using examples from prior attempts. |

Table B3. *The codebook for the content analysis of Q4*

| Category | Definition of category |
|---|---|
| **Most valuable part** | |
| Student-authored | Include when the response values being able to write/enter their own prompts or commands (doing it themselves). |
| Prompt results visualization | Include when the response highlights seeing the prompt output visually (path/simulation/results on screen or phone) as valuable. |
| Scored feedback and | Include when the response values scoring/points/competition or evaluation feedback as valuable. |
| Prompt comparison and | Include when the response values comparing prompts and discussing/explaining why prompts worked or failed (pros/cons, baseline, analysis). |
| Reasoning vs non-reasoning | Include when the response highlights learning differences between reasoning vs non-reasoning models (LLM vs LRM) as valuable. |
| Spot robot | Include when the response identifies the Spot robot (or LLM-controlled robotics) demo as a |
| demonstration | valuable part of the workshop. |
| **Potential improvements** | |
| Clarify system access and map knowledge | Include when the response asks to clarify what the model/system already knows or has access to (e.g., map, obstacles, checkpoints) and what needs to be included in prompts. |
| Clearer task instructions and prompt | Include when the response asks for clearer guidance at the start (what kind of prompts to write, a brief before starting, simpler starter example, or clearer expectations). |
| Pacing and time allocation | Include when the response requests different timing (more time to write prompts, faster/slower review, longer session, less waiting, or less repetitive discussion). |
| More challenging or varied tasks | Include when the response requests harder/more complex tasks, different problems, or different task mix (e.g., more examples, different difficulty balance). |
| More Spot interaction | Include when the response asks for more use of Spot during the workshop (more action, more examples earlier, letting students control, or changes to Spot setup/placement). |
| Logistics, materials, and technical setup | Include when the response requests changes to physical setup or delivery infrastructure (seating, screen visibility, slides/handouts, wifi/QR onboarding, token/API limits, prompt processing speed, class size, or individual device feedback). |
| More conceptual context and real-world relevance | Include when the response asks for more explanation/context (why use these models, prompt-engineering strategies, chain-of-thought concepts, real-world robotics use cases, or broader tradeoffs like environmental impact). |
| **Other impressions** | |
| General positive reaction | Include when the response offers a broad positive evaluation (fun, interesting, engaging, informative, well organized) without a specific detailed theme. |
| No additional feedback | Include when the response explicitly says they have no feedback/changes (e.g., nothing, no comment, no improvements needed). |

### C. Token usage and cost

This workshop collected a total of 494 prompts from consenting participants. Analysis of the total token usage and associated costs for Gemini 2.0 Flash-Lite (Google, 2025a) and Gemini 2.5 Pro (Google, 2025b) revealed an estimated total expenditure of 23.19 USD. This cost was primarily driven by the consumption of thinking tokens (2,221,258 tokens, totaling 22.21 USD). Input token data indicated that the length of students' prompts was highest in Task 1-NR and lowest in Task 2-R, a pattern consistent with the shift from verbose step-by-step instructions for the non-reasoning model to more concise high-level prompts for the reasoning model.

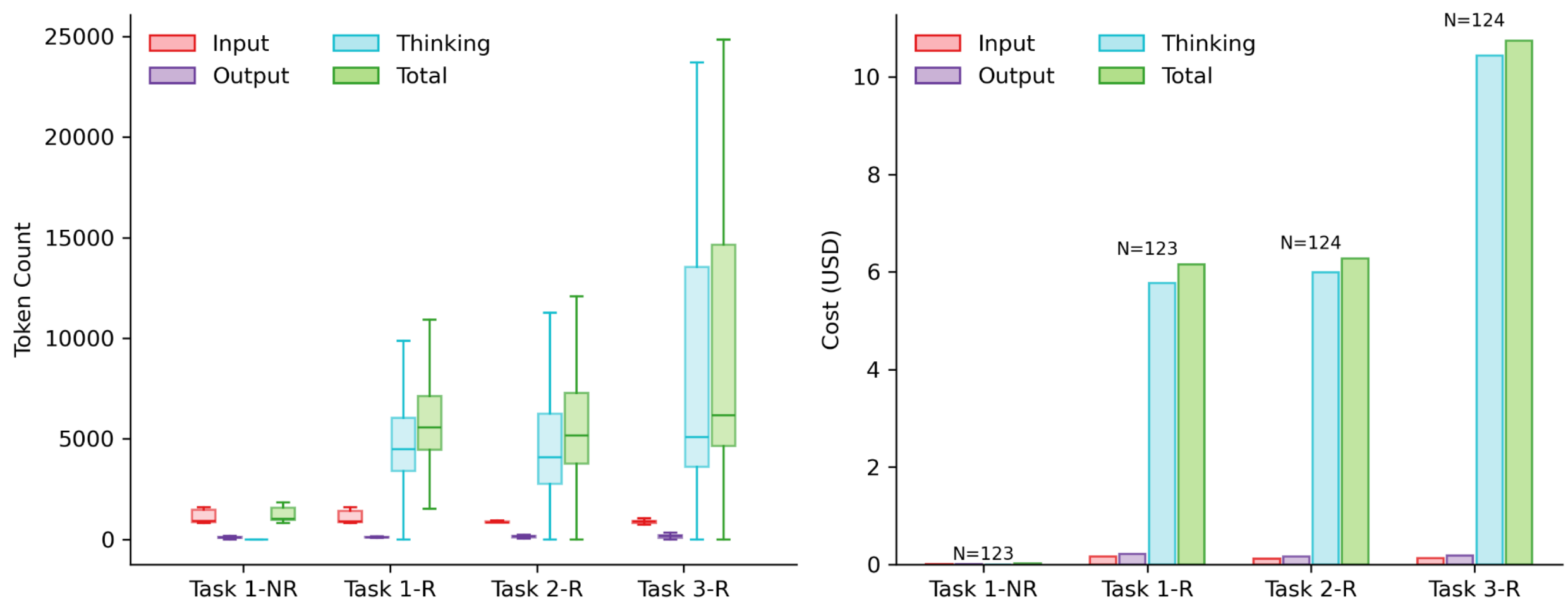

*Figure C1.* Breakdown of token usage and estimated costs.

## D. Approximate workshop timeline

Table D1. *Approximate workshop timeline*

| **Phase** | **Content** | **Duration** |
|---|---|---|
| Introduction | Background knowledge of the workshop | 10 min |
| Interaction | Introduction to Task 1 | 5 min |
| | Writing prompts for Task 1 | 8 min |
| | Running the planner, discussing the outputs | 12 min |
| | Introduction to Task 2 | 2 min |
| | Writing prompts for Task 2 | 4 min |
| | Running the planner, discussing the outputs | 8 min |
| | Introduction to Task 3 | 2 min |
| | Writing prompts for Task 3 | 4 min |
| | Running the planner, discussing the outputs | 8 min |
| Conclusion | Key takeaways, questionnaire | 7 min |
| Demonstration | Real robot navigation & grasping | 10 min |
| Total (not including the Q&A after the demonstration): 80 min | | |